\documentclass{article}

\usepackage[utf8]{inputenc} 
\usepackage[T1]{fontenc}    
\usepackage[margin=1.2in]{geometry}
\usepackage{hyperref}       
\usepackage{url}            
\usepackage{booktabs}       
\usepackage{amsfonts}       
\usepackage{nicefrac}       
\usepackage{microtype}      
\usepackage{lipsum}
\RequirePackage{amsmath, bm}
\usepackage{graphicx}
\usepackage{color}
\usepackage{algorithm}
\usepackage{algpseudocode}
\usepackage{subcaption}
\usepackage{float}
\usepackage{natbib}
\usepackage{hyperref, xr-hyper}
\hypersetup{colorlinks, citecolor=blue, urlcolor=blue}

\DeclareMathOperator*{\argmin}{arg\,min}

\newcommand{\NHPP}{\mbox{NHPP}}

\newcommand{\dd}{\mathrm{d}}

\newcommand{\RI}{\mbox{RI}}

\renewcommand{\hat}{\widehat}
\renewcommand{\bar}{\overline}

\newtheorem{proposition}{Proposition}
\providecommand{\keywords}[1]{\textbf{\textit{Keywords---}} #1}

\title{Heterogeneity Pursuit for Spatial Point
  Pattern with Application to Tree Locations:
  A Bayesian Semiparametric Recourse}

\author{
  Jieying Jiao \\
  Department of Statistics\\
  University of Connecticut\\
  Storrs, CT 06269 \\
  \texttt{jieying.jiao@uconn.edu} \\
   \and
 Guanyu Hu \\
  Department of Statistics\\
  University of Connecticut\\
  Storrs, CT 06269 \\
  \texttt{guanyu.hu@uconn.edu} \\
  \and
  Jun Yan \\
  Department of Statistics\\
  University of Connecticut\\
  Storrs, CT 06269 \\
  \texttt{jun.yan@uconn.edu} \\
}

\begin{document}
\maketitle

\begin{abstract}
Spatial point pattern data are routinely encountered. A
flexible regression model for the underlying intensity is essential to
characterizing the spatial point pattern and understanding the
impacts of potential risk factors on such pattern. We propose a Bayesian
semiparametric regression model where the observed spatial points
follow a spatial Poisson process with an intensity function which
adjusts a nonparametric baseline intensity with multiplicative
covariate effects. The baseline intensity is piecewise constant,
approached with a powered
Chinese restaurant process prior which prevents an unnecessarily large
number of pieces. The parametric regression part allows for variable
selection through the spike-slab prior on the regression coefficients.
An efficient Markov chain Monte Carlo (MCMC) algorithm is developed
for the proposed methods. The performance of the methods is validated
in an extensive simulation study. In application to the locations of
\emph{Beilschmiedia pendula} trees in the Barro Colorado Island forest
dynamics research plot in central Panama,
the spatial heterogeneity is attributed to a subset of soil
measurements in addition to geographic measurements
with a spatially varying baseline intensity.
\end{abstract}


\section{Introduction}\label{sec:intro}
Spatial point pattern data, which are random locations of certain
events of interest in space \citep[e.g.,][]{diggle2013statistical},
arise routinely in many field. Such pattern can be, for example,
locations of basketball shooting attempts in sports analytic
\citep{miller2014factorized,jiao2019bayesian}, earthquake centers in seismology
\citep{schoenberg2003multidimensional}, or tree species in forestry
\citep{leininger2017bayesian, thurman2014variable}. Often times,
spatially varying covariates are available for characterizing the occurrence of
the events. Our motivating application is the spatial point pattern of one of
the most common tree species, \emph{Beilschmiedia pendula}, in the
50--hectare tropical forest dynamics plot at the Barro Colorado Island (BCI)
in central Panama \citep{condit2019complete}. The BCI data is a great
resource for many environmental studies, containing so far complete
information of over $400\,000$ individual trees that have been
censused since the 1980s. It is of interest to characterize the
heterogeneity in the distribution of the tree species by both
spatially varying covariates and a spatially varying baseline
intensity.

Various spatial point process models have been proposed, most of which
allow heterogeneity introduced by covariates. Examples are
Poisson processes \citep{yue2011bayesian}, Gibbs process
\citep{dereudre2019introduction}, pairwise interaction models
\citep{baddeley2000practical}, Neyman--Scott cluster models
\citep{waagepetersen2007estimating}, log-Gaussian Cox processes (LGCP)
\citep{thurman2015regularized, miller2014factorized}, and
modified Thomas processes \citep{yue2015variable},
among others. When likelihood estimation is
infeasible, estimation can be done based on pseudo-likelihood
\citep{baddeley2000practical}. Nonparametric approaches such as kernel
and local likelihood have also been proposed for their flexibility
\citep{baddeley2012nonparametric}. When there are a large number of
covariates, regularized estimation has been proposed
\citep{thurman2014variable, yue2015variable, thurman2015regularized}.

Bayesian approaches have been applied to model spatial point
processes. The empirical Bayes method has long been used, for example,
for inferences for LGCP model parameters \citep{moller1998log}.
Full Bayesian approaches have been proposed for various models such as
Gibbs processes \citep{moller2006efficient, berthelsen2006bayesian,
  king2012bayesian}, LGCPs \citep{moller1998log, illian2012toolbox},
and sequential point processes \citep{moller2012sequential}.
Specifically, Poisson process models are attractive with a rich range
of specifications for the intensity \citep[e.g.,][]{kottas2007bayesian,
  taddy2010autoregressive, yue2011bayesian}. For instance, flexible
nonparametric intensities with a piecewise constant surface can be
modeled by a mixture of finite mixtures \citep{geng2019bayesian}.
Model criticism and model comparison can be based on posterior
predictive samples \citep{leininger2017bayesian}.

The focus of this paper is a Bayesian semiparametric spatial Poisson
point process model which allows nonparametric spatially varying
baseline heterogeneity in addition to covariate-induced heterogeneity.
The nonparametric baseline intensity takes the form of a spatially
piecewise constant function as in \cite{geng2019bayesian}. The
Bayesian framework provide naturally estimates for the number of
components of the piecewise constant function and the component
configuration, along with an estimate of the intensity function
itself. In practice, a widely used Dirichlet process,
namely the Chinese restaurant
process (CRP) \citep{pitman1995exchangeable, neal2000markov} for
nonparametric modeling has been reported to produce overly small
and, hence, redundant components, making the estimator for the number
of components inconsistent \citep{miller2013simple}. To remedy the
situation, \cite{miller2018mixture} put a prior on the number of
componentson, resulting a mixture of finite mixtures model;
\cite{xie2019bayesian} developed a general class of Bayesian
repulsive Gaussian mixture models to encourage well-separated
components. A recent alternative method is the powered
Chinese restaurant process (PCRP) \citep{lu2018reducing}, which
encourages member assignment to existing components. The PCRP method
has not been used in the context of spatial Poisson point process
modeling with a nonparametric baseline as well as covariate effects.

Our contribution is two-fold. First, we propose a flexible Bayesian
semiparametric spatial Poisson point process model. The model
simultaneously captures the heterogeneity introduced by a spatially
varying baseline intensity and the heterogeneity explained by
covariates. The baseline intensity is spatially piecewise constant
with a PCRP prior, which prevents overfitting. Variable selection is
achieved with spike-slab priors \citep{ishwaran2005spike} on the
regression coefficients. The second contribution is an efficient
companion Markov chain Monte Carlo (MCMC) inference. The full
conditional distribution of the index vector of the grid boxes on the
study region under the PCRP prior is summarized in a proposition. The
selection of the power of the PCRP prior is done with a criterion
similar to the Bayesian information criterion. Our
simulation study shows that the proposed method is competitive when
the number of data points in each component is sufficiently large.
Interesting patterns of the \emph{Beilschmiedia pendula} species in
the BCI plot are discovered.

The rest of this paper is organized as follows.
The hierarchical semiparametric model including the spike-slab
prior for variable selection is proposed in Section~\ref{sec:method}.
Details of Bayesian methods for the model such as the MCMC
algorithm, post MCMC inference for the nonparametric baseline, and
selection of the hyperparameter of the PCRP are presented in
Section~\ref{sec:bayes_comp}. A simulation study is reported
in Section~\ref{sec:simu}, followed by an application to the point
pattern of \emph{Beilschmiedia pendula}  from the BIC data in
Section~\ref{sec:real_data}. A discussion concludes in
Section~\ref{sec:discussion}. Additional technical details are
relegated to the Supplementary Materials.

\section{Model Setup}
\label{sec:method}
\subsection{Spatial Poisson Point Process}

The spatial Poisson point process is a fundamental model for spatial
point patterns.  Let
$\mathbf{S} = \{\bm{s}_1, \bm{s}_2, \dots, \bm{s}_N\}$
be the observed points over a study region
$\mathcal{B} \subset \mathbb{R}^2$ with
$\bm{s}_i = (x_i, y_i) \in \mathcal{B}$, $i = 1, \ldots, N$.
A spatial Poisson point process is the process such that the number of
points in any subregion $A\subset \mathcal{B}$ follows a Poisson
distribution with mean
$\lambda(A) = \int_{A}\lambda(\bm{s})\dd \bm{s}$
for some function $\lambda(\cdot)$.
Function $\lambda(\cdot)$ is the intensity function that completely
characterizes the spatial Poisson point process.
When $\lambda(\bm{s})$ changes with $\bm{s}$, the process is known as
a non-homogeneous Poisson Point process (NHPP), denoted by
$\NHPP(\lambda(\cdot))$.

Covariates can be introduced to the intensity function of a NHPP.
Let $\mathbf{X}(\bm{s})$ be a $p\times 1$ spatially varying covariate
vector, which does not include~1. A semiparametric regression model
for the intensity function of an NHPP is
\begin{equation}
  \lambda(\bm{s}_i) = \lambda_0(\bm{s}_i)
  \exp\left(\mathbf{X}^\top(\bm{s}_i)\bm{\beta}\right),
  \label{eq:NHPP}
\end{equation}
where $\lambda_0(\bm{s}_i)$ is an unspecified baseline intensity
function, and $\bm{\beta} = (\beta_1, \dots, \beta_p)^\top$ is a
vector of regression coefficients. The baseline intensity function
$\lambda_0(\cdot)$ captures additional spatial heterogeneity that
are not explained by the covariates.

\subsection{Nonparametric Baseline Intensity}

The baseline intensity function $\lambda_0(\cdot)$ is completely
unspecified in Model~\eqref{eq:NHPP}. For flexibility, we
characterize $\lambda_0(\cdot)$ by a piecewise constant function
\[
  \lambda_0(\bm{s}) = \lambda_{0,z(\bm{s})},
  \qquad z(\bm{s}) \in \{1, \ldots, K\},
\]
where $K$ is the number of components of the piecewise constant
function, vector $\bm{\lambda}_0 = \{\lambda_{0, i}\}_{i = 1}^K$  is
the unique value of $\lambda_0(\bm{s})$, and $z(\bm{s})$ is the index
of the component at location $\bm{s}$. In implementation, we partition
$\mathcal{B}$ by $n$ disjoint grid boxes $A_i$,
$i = 1, 2, \dots, n$, $\mathcal{B} = \cup_{i=1}^nA_i$. Let $z_i$ be
the index of the component in the piecewise constant
function to which $\lambda_0(\bm{s})$ returns for any $\bm{s} \in
A_i$; that is, for any $\bm{s} \in A_i$, we have
$\lambda_0(\bm{s}) = \lambda_{0, z_i}$.

We specify the index process $\bm{z} = (z_1, \ldots, z_n)$ by a PCRP
\citep{lu2018reducing}. In particular, let $z_1 = 1$ and,
for $i \in \{2, \ldots, n\}$,
\begin{equation}\label{eq:pcrp}
\Pr(z_{i} = c \mid z_{1}, \ldots, z_{i-1})  \propto
\begin{cases} {n}_c^r , &  \text{at an existing component labeled}\, c,\\
\alpha, & \text{at a new component},
\end{cases}
\end{equation}
where $\alpha > 0$, $r \ge 1$, and $ n_c$ is the number of
grid boxes in component $c$. This process is denoted by
$\text{PCRP}(\alpha, r)$. The special case of $r = 1$ is the CRP.
When $r > 1$, the PCRP process assigns $z_i$, $i > 1$, to
an existing component with a higher probability than does the CRP
process. This design helps to eliminate artifactual small components
produced by the CRP in modeling mixtures with an unknown
number of components \citep{lu2018reducing}.
As $r$ increases, the probability of each grid box being assigned to
an existing component increases so that the final number of components
needed decreases.

\subsection{Hierarchical Semiparametric Regression Model}

With $\mathbf{S}$ following an $\NHPP$ and index vector $\bm{z}$
following a PCPR, the proposed hierarchical semiparametric regression
model denoted by PCRP-NHPP is
\begin{equation}
  \begin{split}
    \mathbf{S} &\sim \NHPP(\lambda(\bm{s})),\\
    \lambda(\bm{s}) &= \lambda_{0, z_j}
    \exp\left(\mathbf{X}^\top(\bm{s})\bm{\beta}\right),
    \quad \bm{s} \in A_j, \quad j = 1, \ldots, n,\\
    \beta_i &\sim \text{Normal}(0, \delta_i^2),
    \quad i = 1, 2, \dots, p,\\
    \bm{z} &\sim \text{PCRP}(\alpha, r),\\
    \lambda_{0, k} &\sim
    \text{Gamma}(a, b),\quad k = 1, 2, \dots, K,
  \end{split}
  \label{eq:model}
\end{equation}
where $\bm{z} = (z_1, \dots,  z_n)$; $\NHPP(\lambda(\bm{s}))$ and
$\text{PCRP}(\alpha, r)$ are defined in~\eqref{eq:NHPP}
and~\eqref{eq:pcrp}, respectively, with hyperparameters
$(a, b, \delta_1, \ldots, \delta_p, \alpha)$
and a pre-specified power $r$ for
the PCRP; the prior distributions for $\beta_i$'s are independent; the
prior distributions for $\lambda_{0,k}$'s are independent gamma
distributions with mean $a/b$; and $K$ is the number of unique values
of $z_i$.

When variable selection is desired,
a spike-slab prior can be imposed on each element of $\bm{\beta}$
\citep{ishwaran2005spike}. A spike-slab distribution is a mixture of
a nearly degenerated distribution at zero (the spike) and a flat
distribution (the slab). Zero-mean normal distributions with a small
and a large variance are common choices, respectively, for the spike
and the slab. The ratio of the two variances should be in a reasonable
range such that the MCMC will not get stuck in the spike
component \citep{malsiner2018comparing}. Following the suggestion
from \cite{george1993variable}, a common choice of the two variances
are 0.01 and 100. Specifically, the normal spike-slab prior specifies
the hyperparameter $\delta_i$, $i = 1, \ldots, p$,
in~\eqref{eq:model} as
\begin{align}
  \begin{split}
    \delta^2_i &= 0.01(1-\gamma_i) + 100\gamma_i,\\
    \gamma_i &\sim \text{Bernoulli}(0.5).
  \end{split}
  \label{eq:spikeslab}
\end{align}
With posterior samples from MCMC, variable selection is done using the
posterior modes of~$\gamma_i$'s. A posterior mode zero of $\gamma_i$
suggests little
significance of $\beta_i$ and exclusion of the corresponding covariate
from the regression model.

\section{Bayesian Inference}
\label{sec:bayes_comp}

\subsection{The MCMC Sampling Scheme}
The parameters in Model~\eqref{eq:model}--\eqref{eq:spikeslab} are
$\bm{\Theta} = \{\bm{\lambda}_0, \bm{z}, \bm{\beta}, \bm{\gamma}\}$,
where $\bm{\gamma} = (\gamma_1, \dots, \gamma_p)$. The likelihood of
spatial Poisson point process is
\begin{equation*}
  L(\bm{\Theta}  | \mathbf{S}) = 
  \prod_{i=1}^N \lambda(\bm{s}_i)
  \exp\left(-\int_{\mathcal{B}}\lambda(\bm{s}) \dd \bm{s}\right).
\end{equation*}
The posterior density of $\bm{\Theta}$ is
\[
  \pi(\bm{\Theta} | \textbf{S}) \propto
  L(\bm{\Theta} | \mathbf{S}) \pi(\bm{\Theta})
\]
where $\pi(\bm{\Theta})$ is the prior density of $\bm{\Theta}$.

The full conditional distribution of each parameter in $\bm{\Theta}$
are derived in order to use Gibbs sampling method; see
Section A in the Appendix.
The full conditional distribution for the elements in the index
process $\bm{z}$ is summarized in the following Proposition.
\begin{proposition} \label{thm:z}
Under the model and prior specification~\eqref{eq:model}, the full
conditional distribution of $z_i$, $i = 1, \ldots, n$, is
\begin{equation}
  \Pr(z_i = c\mid \mathbf{S}, \bm{z}_{-i},\bm{\lambda}_0, \bm{\beta})
  \propto
  \begin{cases}
    n_{-i, c}^r\lambda_{0, c}^{m_i}
    \exp(- \lambda_{0, c}\Lambda_i(\bm{\beta}))
    & \exists j \ne i, \, z_j = c \quad \mbox{(existing label)},\\
    \displaystyle{
      \frac{\alpha b^a \Gamma(m_i+a)}
      {(b+\Lambda_i(\bm{\beta}))^{m_i+a} \Gamma(a)}
    }
    & \forall j \ne i, \, z_j \ne c \quad \mbox{(new label)},
  \end{cases}
  \label{eq:z}
\end{equation}
where $\bm{z}_{-i}$ is $\bm{z}$ with $z_i$ removed,
${n}_{-i, c}$ is the number of grid boxes in component $c$ excluding
$A_i$, $m_i = \sum_{j = 1}^N 1(\bm{s}_j \in A_i)$ is the number of data
points in grid box $A_i$, and $\Lambda_i(\bm{\beta}) =
\int_{A_i}\exp\big(\mathbf{X}^\top(\bm{s})\bm{\beta}\big)\dd \bm{s}$,
The results remain when the spike-slab prior is imposed on the 
elements of $\bm{\beta}$. 
\end{proposition}

Assuming that the covariates $\mathbf{X}(s)$ are piecewise constant on
a grid partition, which may not be the same as the $A_i$'s, the
integral of $\Lambda_i(\bm{\beta})$ in Proposition~\ref{thm:z}
can be calculated as a summation on each $A_i$, $i = 1, \ldots, n$.
If this grid partition is the same as $A_i$'s,  that is,
$\mathbf{X}(\bm{s}) = \mathbf{X}_i$ for $\bm{s}\in A_i$, then
$\Lambda_i(\bm{\beta}) = \mu(A_i)
\exp(\mathbf{X}_i^\top\bm{\beta})$,
where $\mu(A_i)$ is the area of $A_i$.

The full conditional distributions for $\lambda_{0,k}$'s and
$\gamma_j$'s are straightforward from their conjugate priors:
\begin{align}
  \lambda_{0, k} \mid
  \mathbf{S}, \bm{\beta}, \bm{\gamma}, \bm{z}, \bm{\lambda}_{0, -k}
  &\sim
    \text{Gamma}
    (N_k+a,
    b+\sum_{j:z_j = k}\Lambda_j(\bm{\beta})),
    \quad k = 1, \ldots, K,
    \label{eq:lambda}\\
  \gamma_j \mid
  \mathbf{S}, \bm{\beta}, \bm{\gamma}_{-j},\bm{z}, \bm{\lambda}_0
  &\sim
    \text{Bernoulli}
    \left(\left(1+
    \frac{\phi(\beta_j | 0.01)} {\phi(\beta_j |100)}\right)^{-1}
    \right),    
    \quad j = 1, \ldots, p,
    \label{eq:gamma}
\end{align}
where
$\bm{\lambda}_{0, -i}$ and $\bm{\gamma}_{-i}$ are, respectively
the $\bm{\lambda}_{0}$ and $\bm{\gamma}$ without the $i$th element,
$N_k = \sum_{i: z_i = k}m_i$ is the number
of data points in the $k_{th}$ component, and $\phi(\cdot | \sigma^2)$
is the density of $\text{Normal}(0, \sigma^2)$.

The full conditional density of $\beta_j$,
$q(\beta_j \mid \mathbf{S}, \bm{\beta}_{-j}, \bm{\gamma},
    \bm{z}, \bm{\lambda}_0)$  is proportional to
\begin{equation}
  \begin{split}
 \phi^{1 - \gamma_i} (\beta_j | 0.01) \phi^{\gamma_i} (\beta_j | 100)
 \prod_{i=1}^n {\lambda}_{0, z_i}^{m_i}
    \exp \left(\sum_{\ell:\bm{s}_\ell \in A_i}
      \mathbf{X}^\top(\bm{s}_\ell)\bm{\beta} - \lambda_{0,
        z_i}\Lambda_i(\bm{\beta}) \right), \quad j = 1, \ldots, p,
  \end{split}
  \label{eq:beta}
\end{equation}
where $\bm{\beta}_{-i}$ is $\bm{\beta}$ without the $i$th element. The
Metropolis--Hastings algorithm \citep{hastings1970monte}
can be used to draw samples from
this conditional distribution. In our implementation, we used a normal
distribution proposal centered at the current value with a standard
deviation that is tuned to achieve a desired acceptance rate.

The full conditional distributions facilitate MCMC sampling with an
Gibbs sampling algorithm. Algorithm~\ref{alg:mcmc} summarizes the
specifics for one MCMC iteration. In the algorithm, $K$, the length of
$\bm{\lambda}_0$, reduces by~1 whenever a component is found to
contain only a single grid box; it increases by~1 when a new label is
assigned by the full conditional distribution~\eqref{eq:z}.
To initialize, each $z_i$, $i = 1, \ldots, n$, is randomly assigned an
integer value in $\{1, \ldots, K_0\}$ for a prespecified initial
number of components $K_0$ for the piecewise constant
baseline, possibly based on some exploratory analysis. Initial values
for $\bm{\lambda}_0$ are generated independently from the
$\mathrm{Gamma}(a, b)$ prior distribution. Initial values for other
parameters $\bm{\beta}$ and $\bm{\gamma}$ can simply be set to zeros.

\begin{algorithm}[tbp]
  \caption{Gibbs sampling algorithm for one iteration of MCMC to
    update $\bm{\Theta}$.}
  \label{alg:mcmc}
  \begin{algorithmic}[1]
  \State update $\Lambda_i(\bm{\beta})$, $i = 1, \ldots, n$
  \For{$i = 1:n$} \Comment{Update $\bm{z}$}
  \If{$A_i$ is the only grid box in the component that it belongs to}
  \State $K = K-1$
  \State $z_j = z_j - 1$ for all $z_j$ such that $z_j > z_i$
  \State shorten $\bm{\lambda}_0$ by dropping $\lambda_{0, z_i}$
  \EndIf
  \State draw $z_i$ from~\eqref{eq:z}
  \If{$z_i$ goes to a new component}
  \State $K = K+1$
  \State draw $\lambda_{0, K} \sim \mbox{Gamma}(a, b)$
  \EndIf
  \EndFor
  \For{$i = 1:K$} \Comment{Update $\bm{\lambda}_0$}
  \State draw $\lambda_{0, i}$ from~\eqref{eq:lambda}
  \EndFor
  \For{$i = 1:p$} \Comment{Update $\bm{\gamma}$}
  \State draw $\gamma_i$ from~\eqref{eq:gamma}
  \EndFor
  \For{$i = 1:p$} \Comment{Update $\bm{\beta}$}
  \State draw $\beta_i$ from~\eqref{eq:beta} with the
  Metropolis--Hastings algorithm
  \EndFor
\end{algorithmic}
\end{algorithm}

\subsection{Inference for the Nonparametric Baseline Intensity}
\label{sec:sum_mcmc}

Bayesian inference for the nonparametric baseline intensity is not as
straightforward as for the regression coefficients whose posterior
sample can be easily summarized. The nonparametric baseline intensity
is constructed based on the index vector $\bm{z}$, which can not be
summarized using traditional posterior mean or mode. The same $z_i$
value from different iterations does not necessarily mean the same
component. The similar problem also exists in inference of
$\bm{\lambda}_0$. We use Dahl's method \citep{dahl2006model} as a
simple solution for summarizing $\bm{z}$. It chooses the iteration in
the posterior sample that optimizes a least squares criterion as the
estimate for $\bm{z}$.

For a draw from the posterior distribution of $\bm{\Theta}$, define
an $n \times n$ membership matrix
\begin{align}\label{eq:membermat}
  B = (B(i,j)) =
  \big( 1(z_i = z_j)\big), \quad i,j\in \{1,2,\dots,n\}.
\end{align}
That is, its $(i,j)$-th element is~1 if the $i$-th and $j$-th grid boxes
belong to the same component (or have the same baseline intensity); it
is~0 otherwise. For an MCMC sample of size $M$, let $B^{(t)}$ be the
$B$ matrix defined for the $t$th draw in the sample and
$\bar{B} = \frac{1}{M} \sum_{t=1}^{M} B^{(t)}$. Then each $(i, j)$-th
element of $\bar B$ is the relative frequency that the $i$-th and $j$-th
grid boxes belong to the same component. \cite{dahl2006model}
suggested to take the draw in the sample that is closest to~$\bar{B}$
as the point estimate of $\bm{z}$. Let
\begin{equation*}
    t_* =  \argmin_{t \in \{1,2,\dots, M\}} \sum_{i=1}^n \sum_{j=1}^n
             (B^{(t)}(i,j) - \bar{B}(i,j))^2.
\end{equation*}
The estimate for $\bm{\lambda}_0$ is $\bm{\lambda}^{(t_*)}_0$
and the estimate of $K$ is the length of~$\bm{\lambda}^{(t_*)}_0s$.
The advantage of this method is that is uses
information from all posterior samples, and the final result is
guaranteed to be a valid scheme that exists in the sample.

Convergence check for the nonparametric baseline cannot be done with
trace plots for $\bm{z}$ or $\bm{\lambda}_0$ as their meanings change
from iteration to iteration. The baseline at each grid box,
$\lambda_{0, z_j}$, $j = 1, \ldots, n$, does have the same meaning
across iteration and can be checked for convergence. The situation is
similar to that of a reversible jump MCMC where a nonparametric
component has varying degrees of freedom \citep{wang2013bayesian}.
Instead of monitoring a large number $n$ grid boxes, we focus on the
number of component $K$ in practice and monitor the trace plot of $K$
for stationarity. Usually, once $K$ shows stationarity, the
nonparametric baseline at each grid box and the regression
coefficients are all stationary.

As a further diagnostic tool for the nonparametric baseline, the Rand
index (RI), which measures the similarity of two component memberships
\citep{rand1971objective}, can be checked. Let
$\Omega = \{(A_i, A_j)\}_{1 \le i < j  \le n}$
be the collection of all pairs of grid boxes.
For two index vectors $\bm{z}^{(1)}$ and $\bm{z}^{(2)}$,  define
\begin{equation*}
  \begin{split}
    a &= \# \{(A_i, A_j) \in \Omega : z_i^{(1)} =
        z_j^{(1)}, z_i^{(2)} = z_j^{(2)}\},\\
    b &= \# \{(A_i, A_j) \in \Omega : z_i^{(1)} \ne
        z_j^{(1)}, z_i^{(2)} \ne z_j^{(2)}\},
  \end{split}
\end{equation*}
where $\#\{\cdot \}$ denotes the cardinality of a set. That is, under
the two membership assignments, $a$ and $b$ are the number of grid box
pairs whose memberships are ``concordant''; other pairs are all
``discordant''. Then, RI is defined as
\[
  \mbox{RI} 
  ={n \choose 2}^{-1} (a + b),
\]
which ranges from 0 to 1 with a higher value indicating a better
agreement between the two index vectors, and~1 indicates a perfect
match.

If the true index vector $\bm{z}_0$ were known, for
every MCMC iteration~$\bm{z}^{(t)}$, $t = 1, 2, \dots, M$, a
Rand index $\RI^{(t)}$ based on $\bm{z}_0$ and $\bm{z}^{(t)}$ can be
calculated. The trace plot of $\RI^{(t)}$ is a useful diagnostic tool.
In practice, since $\bm{z}_0$ is unknown, we replace it with the
estimate $\hat{\bm{z}}$ from Dahl's method after convergence has been
ensured from the trace plot of $K$. The stationarity in the trace plot
of RI provides a second stage reassurance of the convergence, and its
level provides a measure for the quality of the consistency of the
memberships from iteration to iteration.

\subsection{Selection of $r$}
The estimated number of components $\hat K$ of the nonparametric
baseline intensity depends on $r$, the power of the PCRP prior.
Selection of~$r$ remains to be addressed. Model comparison
criteria under the Bayesian framework, such as deviance information
criterion (DIC) \citep{spiegelhalter2002bayesian} and logarithm
of pseudo-marginal likelihood (LPML) \citep{geisser1979predictive,
  gelfand1994bayesian}, are natural choices. From our simulation
study, however, they tended to lead to overestimation of the number of
components $K$ in the nonparametric baseline. We experimented with a
criterion in the spirit of the Bayesian information criterion (BIC)
since BIC has proven to be an effective criterion for likelihood-based
model selection in clustering algorithms~\citep{wang2017likelihood}.

Our Bayesian information type criterion (BITC) is defined as
\begin{equation}
  \mbox{BITC} = -2\log L(\hat{\bm{\Theta}} \mid \mathbf{S}) + \hat{K}\log(N),
  \label{eq:BIC}
\end{equation}
where $\hat{\bm{\Theta}}$ and $\hat{K}$ are estimator of
$\bm{\Theta}$ and $K$, respectively, from Dahl's method.
Although BIC is usually used for frequentist methods, when number of
points $N$ and the MCMC sample size $M$ are large enough,
$\hat{\bm{\Theta}}$ provides a consistent estimator of $\bm{\Theta}$
\citep{walker1969asymptotic} and it is reasonable to use BITC to
assess model fitting without considering parameter variation. The
effectiveness of the BITC in selecting $r$ was confirmed in our
simulation study reported in the next section. In practice, we can
determine the range of candidate $r$ values starting from~1 and ending
with a number that gives $\hat K = 1$. Then the optimal $r$ is
selected based on the BITC from a grid of candidate $r$ values. The
optimal $r$ selected by this criterion provided better estimate of $K$
in our simulation than those by LPML and DIC. Its performance in
general model comparison is beyond the scope of this paper.

\section{Simulation Study}\label{sec:simu}

The proposed methods were validated in a simulation study over a
region $\mathcal{B} = [0, 20]\times [0, 20]$. The study region is
partitioned by grid boxes $A_i$, $i = 1, 2, \dots, n$, where each
$A_i$, $i \in \{1, 2, \dots, n = 400\}$, is a unit square.
Points were generated from the $\NHPP(\lambda(\bm{s}))$
model~\eqref{eq:NHPP} with $p = 4$ covariates.
The four covariates $X_i(s)$, $i = 1, 2, 3, 4$,  were set to be
piecewise constant over the grid boxes $A_i$'s. Their values were
independently generated from the standard normal distribution.
The true regression coefficients were $\beta_1 = \beta_2 = 0.5$ and
$\beta_3 = \beta_4 = 0$. Two settings of the piecewise constant
baseline intensity surface were considered. 
Setting~1 had two components, with
$\bm{\lambda}_0 = (0.2, 10)$, and the number of grid boxes in the two
components were $(n_1, n_2) = (309, 91)$.
Setting~2 had three components, with
$\bm{\lambda}_0 = (0.2, 5, 20)$, and
$(n_1, n_2, n_3) = (232, 91, 77)$.
See Figure~\ref{wfig:baseline} for the spatial structure of the
baseline intensity surfaces under the two settings. In both settings,
more grid boxes were assigned to first component with low
intensity value, which mimics the pattern of the BCI data. We used
function \texttt{rpoispp()} from \textsf{R} package
\textsf{spatstat} \citep{baddeley2015spatial} to generate the points.
For each setting, 100 replicates were generated. There were about
$1000$ to $1500$ data points generated under setting~1, and
about $3000$ to $4000$ data points generated under setting~2.

Priors for the model parameters were set to be
those in~\eqref{eq:model} and~\eqref{eq:spikeslab}, with
hyperparameters $a = b = \alpha = 1$.
Different $r$ values starting from~1 were tried on each dataset and
the optimal $r$ was selected by the BITC in~\eqref{eq:BIC}. The upper
end of the candidate $r$ in each setting was experimented to be
the smallest value that would lead to a single component of the
piecewise constant baseline intensity surface based on Dahl's method.
Candidate powers used for setting 1 were from 1 to 2 with a step 0.1;
for setting 2, candidate
powers used were from 1 to 3 with a step 0.1.
In the Metropolis--Hastings algorithm to draw $\bm{\beta}$,
a Normal$(0, 0.05^2)$ distribution was used as the proposal, which
yielded an acceptance rate of 30--40\%. For each dataset, we ran the
MCMC for 5000 iterations and discarded the first 1000 iterations.
Convergence for the remaining iterations was ensured by checking
the trace plots of the elements in $\bm{\beta}$ and $K$.

\begin{figure}[tbp]
  \centering
  \subcaptionbox{Histograms of K}{
    \includegraphics[width=\textwidth]{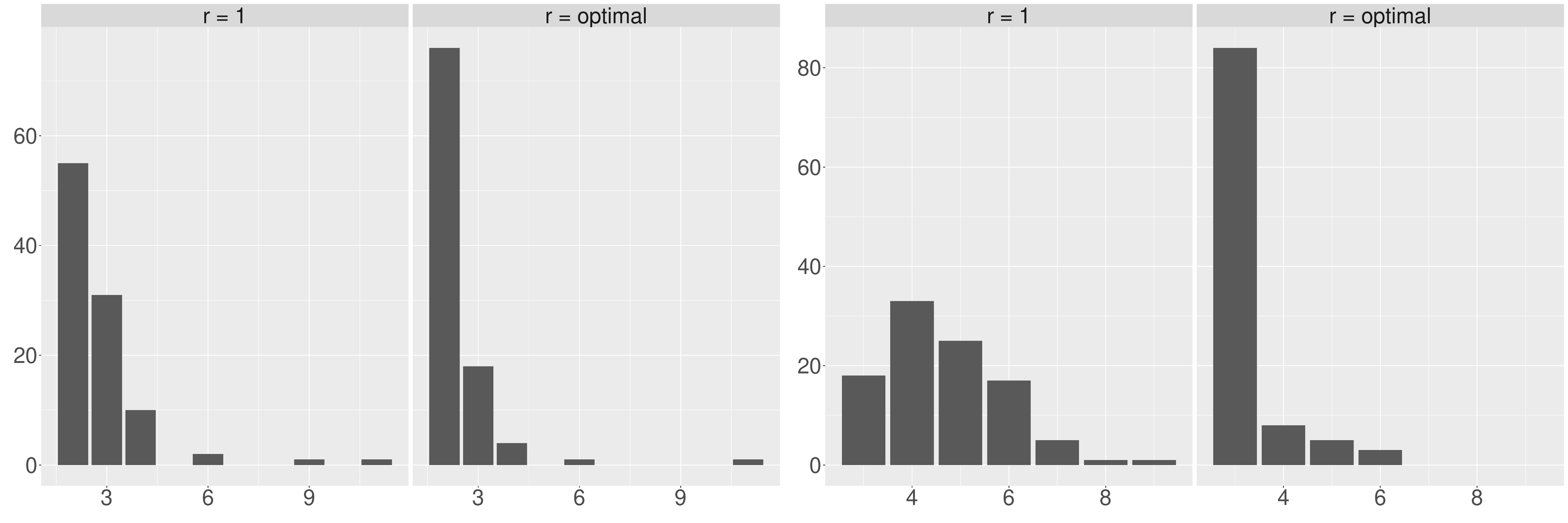}}
  \subcaptionbox{Rand Index}{
    \includegraphics[width=\textwidth]{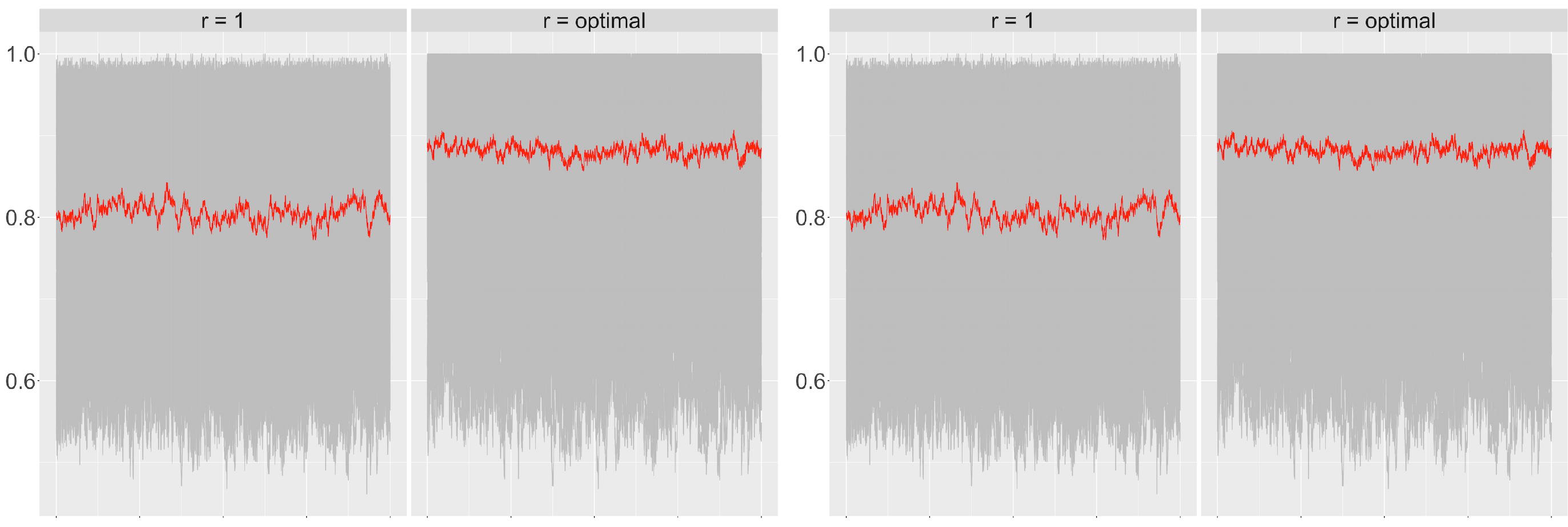}}
  \caption{Histograms of $\hat K$ and overlaid trace plots of RI from
    the 100 replicates in simulation study. Setting 1 and 2 are on
    left and right panel, respectively. The optimal $r$ was selected
    by the BITC. The thick lines are the average of the trace plots
    over the 100 replicates.}
  \label{fig:K_RI}
\end{figure}

Figure~\ref{fig:K_RI} shows the histograms of $\hat K$ from
Dahl's method for a selective set of $r$ values based on the 100
replicates. The optimal~$r$ was selected by the BITC. The true $K$
are~2 and~3, respectively, in the two settings. The histograms of
$\hat K$ under the optimal~$r$ are much closer to the true $K$ under
both settings than those under $r = 1$, the popular CRP prior.
Among the 100 replicates, there are 76 times in setting~1 and 84
times in setting~2 that the optimal $r$ led to $\hat K = K$ exactly.
The same frequencies for $r = 1$ are only 55 and 18,
respectively. LPML and DIC were also calculated using the Monte Carlo
estimation proposed by~\cite{hu2019new}. As shown in
Figure~\ref{wfig:K_hist2}, they had good
performance under setting~1, estimating $K$ correctly for 81 times and
83 times, respectively. Nonetheless, both of them did fairly poorly
under setting 2 with obviously more components estimated than the
truth. Only 24 times and 23 times out of the 100 replicates estimated
$K$ correctly using LPML and DIC, respectively.

Also shown in Figure~\ref{fig:K_RI} are the overlaid trace plots of
the RI from all the replicates under $r = 1$ and the optimal $r$.
The trace plots further assure the convergence of the index process
$\bm{z}$ in the sense of \cite{rand1971objective}. The averaged trace
plots over the replicates (shown in solid lines) reflect the level of
consistency in component memberships across iterations. Under
$r = 1$, the averages stabilize around 0.806 and 0.828, respectively,
for setting~1 and setting~2. Under the optimal $r$, the averages are
0.882 and 0.893, respectively, indicating higher consistencies in
component membership than those under $r = 1$.

The estimated baseline intensity surfaces under the optimal $r$ for the
100 replicates are summarized, and the heat maps of their $2.5\%$
percentile, median and $97.5\%$ percentile are compared with the
true surface in Figure~\ref{wfig:baseline}. According to the
plots, fitted baseline
intensity can accurately specify different components, and the
intensity magnitude is also very close to the truth. The estimation of
the edge part of different components do not show obvious worse
performance. This is because PCRP does not have spatial smoothness
constraint and allows more spatial flexibility. As long as the
resolution of grid box is fine enough, PCRP can capture different
components no matter the shape of the area. Under each setting, we
pool the grid boxes under the same component 
and compare the averaged estimates with the true value.
For setting~1, the empirical biases for the two components of
$\bm{\lambda}_0 = (0.2, 10)$ are $(0.052, -0.413)$, with standard
deviations $(0.055, 0.433)$.
For setting~2, the empirical biases for the three components
$\bm{\lambda}_0 = (0.2, 5, 20)$ are $(0.089, 0.107, -0.846)$,
with standard deviations $(0.058, 0.381, 0.752)$.
Considering the magnitude of the intensity components, they are
estimated quite accurately.

\begin{table}
  \caption{Simulation estimation results. ``SD'' is the empirical standard
    deviation over 100 replicates, ``$\widehat{\mathrm{SD}}$'' is the
    average of 100 standard deviations calculated using posterior
    sample. ``AR'' is the variable selection accuracy rate over 100
    replicates.}
  \label{tab:simu_est}
  \centering
  \begin{tabular}{cccccc}
    \toprule
    Setting & $\beta$ & AR($\%$)
    & Bias & SD & $\widehat{\mathrm{SD}}$ \\
    \midrule
    Setting 1 & 0.5 & 100 & $\phantom{-}$0.019 & 0.035 & 0.033\\
    &0.5 & 100 & $\phantom{-}$0.016 & 0.032 & 0.033\\
    &0.0 & 100 & $\phantom{-}$0.028 & 0.034 & 0.030\\
    &0.0 & 100 & $\phantom{-}$0.022 & 0.035 & 0.031\\
    \\
    Setting 2 &0.5 & 100 & $\phantom{-}$0.012 & 0.027 & 0.026\\
    &0.5 & 100 & $\phantom{-}$0.011 & 0.025 & 0.025\\
    &0.0 & 100 & $\phantom{-}$0.007 & 0.024 & 0.023\\
    &0.0 & 100 & $\phantom{-}$0.003 & 0.021 & 0.023\\
    \bottomrule
  \end{tabular}
\end{table}

Table~\ref{tab:simu_est} summarizes the results of variable selection
and estimation for the parametric regression part of the model.
For both settings, the two important variables were included and the
two unimportant variables were excluded with accuracy rate 100\%.
The empirical bias of the coefficients are minimal. The empirical
standard errors of the point estimates have good agreement with the
average of the posterior standard deviation of the coefficients. The
variation of the coefficient estimates is lower in setting~2 than in
setting~1, which is expected because setting~2 has high intensity and
more observed points.

Finally, the proposed model was compared with three competing models. 
The first one, denoted as ``const-NHPP'', is the
$\NHPP(\lambda(\bm{s}))$ model~\eqref{eq:NHPP} with a
constant baseline intensity surface $\lambda_0$.
The second one, denoted as ``spline-NHPP'', is the
$\NHPP(\lambda(\bm{s}))$ model~\eqref{eq:NHPP} with baseline intensity
surface approximated by a tensor product cubic splines with a single
knot for each of the two coordinates \citep[e.g.,][]{berhane2008using}.
This model can be fit with the spline basis into the covariates. 
The last model is the LGCP model fitted with function \texttt{kppm()}
from \textsf{R} package \textsf{spatstat} \citep{baddeley2015spatial}.
The mean squared error (MSE) is often used to compare the performances
of different models in modeling spatial point pattern, which, with a
grid partition $A_i$'s on $\mathcal{B}$, is defined as
\begin{equation}
\text{MSE}=\frac{1}{n}\sum_{i=1}^n\left|\mu(A_i)\hat{\lambda}(A_i)-m_i
\right|^2,
\label{eq:MSE}
\end{equation}
where
$\hat{\lambda}(A_i) =
\int_{\bm{s}\in A_i} \hat{\lambda}(\bm{s})\dd \bm{s}$,
$\hat{\lambda}(\bm{s})$ is the estimated 
intensity on location $\bm{s}$. Models with
smaller MSE values are preferred.

\begin{figure}[tbp]
  \centering
  \includegraphics[width=\textwidth]{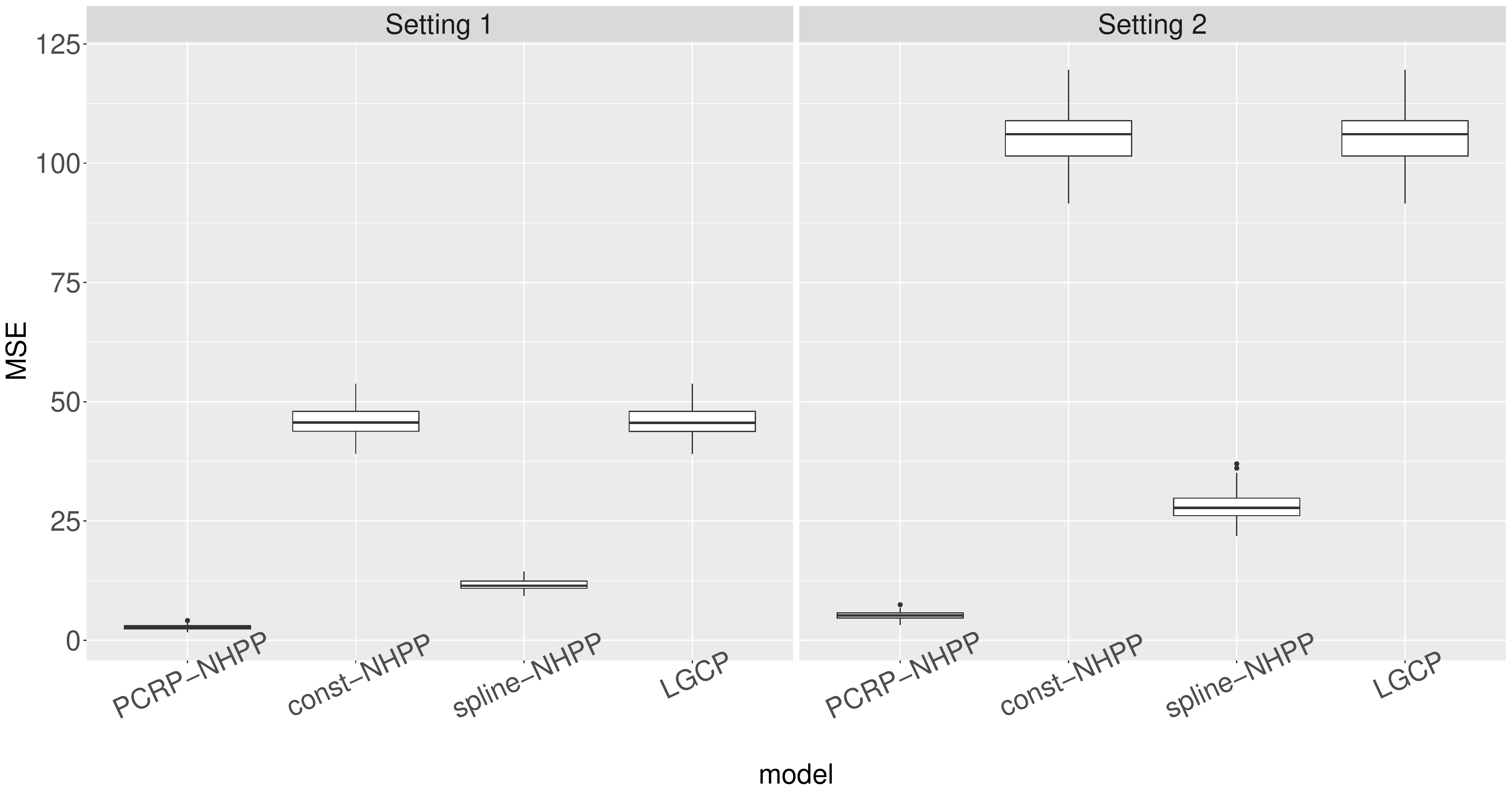}
  \caption{Boxplots of MSE over the 100 replicates for four models in
    simulation study.}
  \label{fig:simu_MSEbox}
\end{figure}

Figure~\ref{fig:simu_MSEbox} summarizes the boxplots of the MSE
across the proposed model and the three competing models.
The three competing models are all misspecified under the data
generating schemes. As expected, they have much higher MSEs than the
proposed model. In particular, the const-NHPP and LGCP models show
very similar results, since the LGCP model mainly makes improvements
on the intensity's variance structure, which does not help here.
The MSE of the spline-NHPP model could be further reduced if more
flexibility in the splines were introduced. Nonetheless, its baseline
intensity is spatially continuous which may never fit well the data
generated in our settings with piecewise constant intensity surface
consisting of only two or three quite different components.

\section{Point Pattern of \emph{Beilschmiedia
    Pendula}}\label{sec:real_data}

\begin{figure}[tbp]
  \centering
  \includegraphics[width=\textwidth]{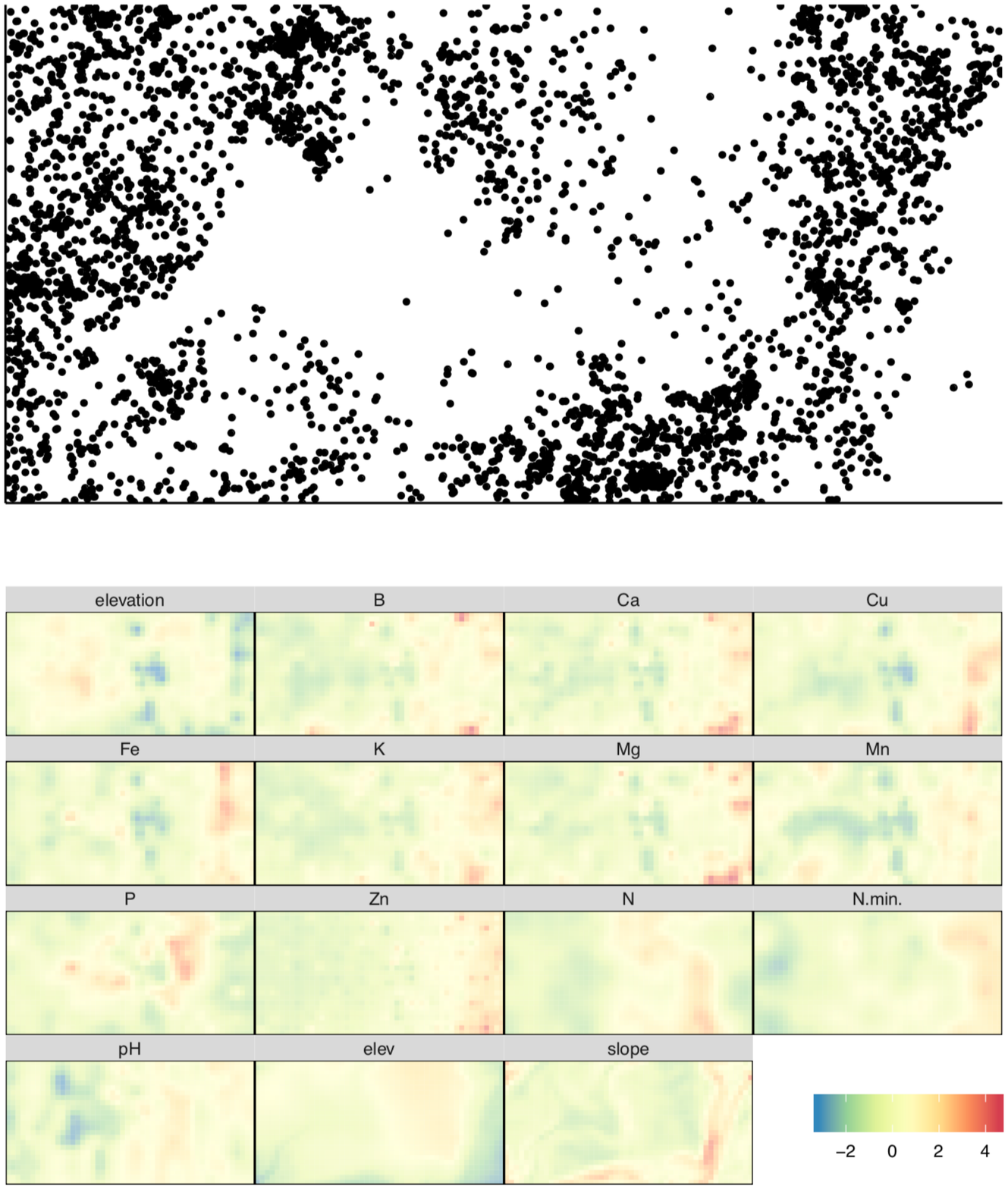}
  \caption{
  The locations of \emph{Beilschmiedia Pendula} and heat maps of
  the standardized covariates of the BCI data.}
  \label{fig:dataplot}
\end{figure}

\emph{Beilschmiedia pendula} is one of the most abundant tree species
in the BCI \citep{thurman2014variable}. There are $4,194$ such trees
in total from the most recent census in the 50-hectare plot $\mathcal{B}$,
a rectangle of $1000m\times 500m$. Their exact locations are recorded
in a Cartesian coordinate system
$(x, y) \in \mathcal{B} = [0, 1000]\times[0, 500]$;
see Figure~\ref{fig:dataplot}. In addition to two
geographical variables, elevation and slope, thirteen environmental
covariates are available, which are soil pH and concentrations in the
soil of aluminum (Al), boron (B), calcium (Ca),
cuprum (Cu), ferrum (Fe), kalium (K), magnesium (Mg), manganese (Mn),
phosphorus (P), zinc (Z), nitrogen (N), and nitrogen
mineralization (N.min.). The latest version of these covariates data
was from 2004, available at
~\url{http://ctfs.si.edu/webatlas/datasets/bci/soilmaps/BCIsoil.html},
These variables were measured at each of the
$1,250$ grid boxes of size $20m\times 20m$ over
$\mathcal{B}$. Within each grid box, the measures are assumed to be
the same so that each measure is piecewise constant over $\mathcal{B}$.
The heat maps of the standardized covariates are also shown in
Figure~\ref{fig:dataplot}.

With standardized covariates, the hierarchical semiparametric
model~\eqref{eq:model}--\eqref{eq:spikeslab} were fit to the observed
point pattern of \emph{Beilschmiedia Pendula}. For better numerical
performance, the area of each grid box was scaled to be~1. The
hyperparameters were set to be $a = b = \alpha = 1$.
The standard deviation of the normal proposal in the
Metropolis--Hastings algorithm was set to be 0.05 in order to make the
acceptance rate between $30\%$ and $40\%$.
A grid of values $\{1, 1.1, 1.2, 1.3, 1.4, 1.5\}$ were used for $r$ in
search for an optimal $r$. For each $r$, an MCMC was carried out for
50,000 iteration. The first 10,000 iterations dropped as burnin, and
the remaining iterations were thinned by 10, yielding an MCMC sample
of size $M = 2000$. The convergence was checked for the trace plots of
$\bm{\beta}$ and $K$; see Figure~\ref{wfig:trace}.

\begin{figure}[tbp]
  \centering
  \includegraphics[width=\textwidth]{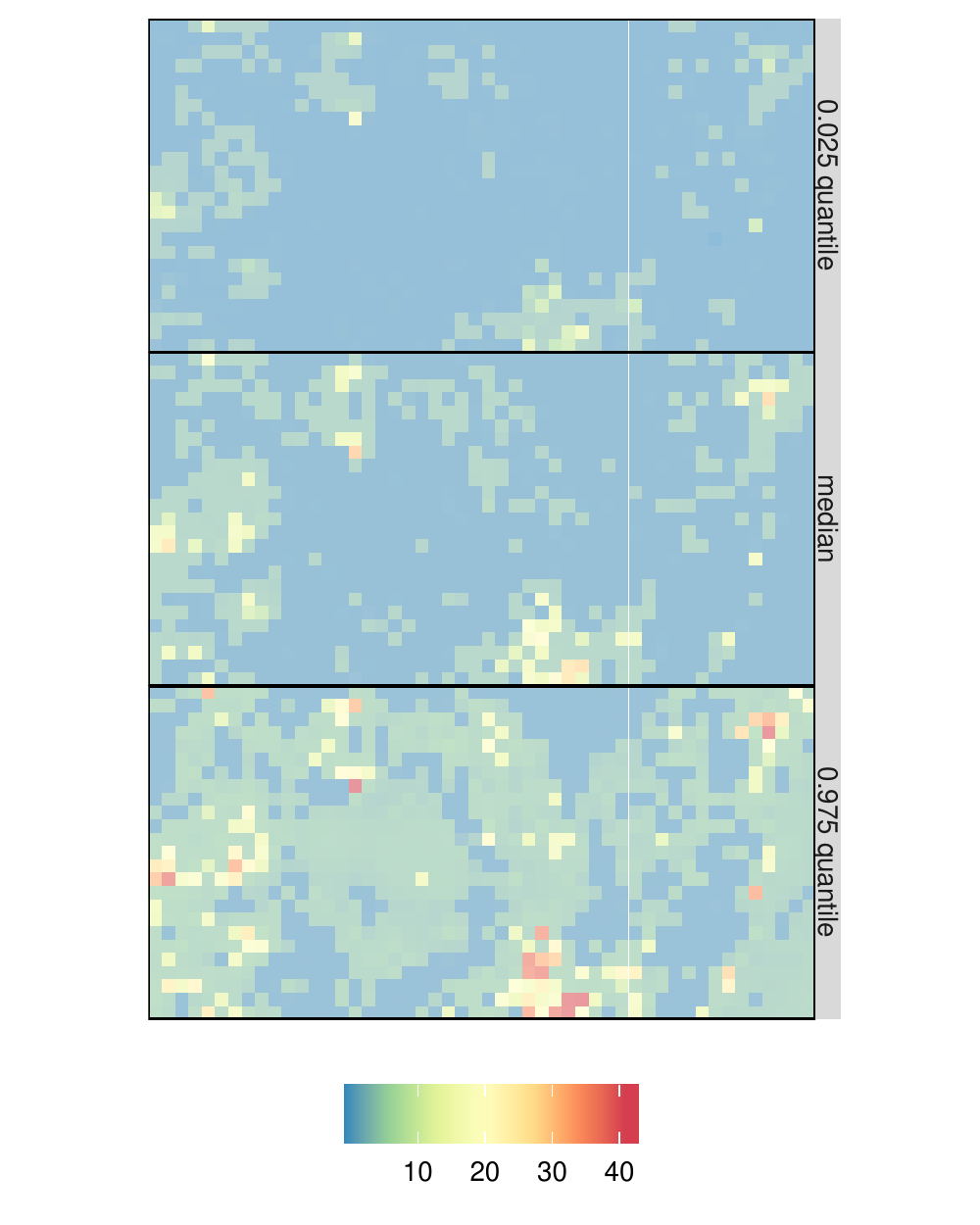}
  \caption{Heat map of the fitted baseline intensity surface of the
    BCI data using $r = 1.3$. From top to bottom: 2.5\%, 50\%, and
    97.5\% percentiles of posterior distribution of $\lambda(\bm{s})$.}
  \label{fig:Realdata_intensity}
\end{figure}

The optimal $r$ was selected to be 1.3 by the BITC, which leads to
$\hat K = 4$ from Dahl's method. The posterior mode of $K$ is
also~4, with a frequency of 1198 out of 2000. The posterior standard
deviation of $K$ is 0.852.  About 93.4\% of the sample values of $K$
are $\{3, 4, 5\}$, which is rather tight compared to those from the
CRP prior for the baseline ($r = 1$). The average RI over
the MCMC sample was 0.806, suggesting good concordance of the component
assignments over the MCMC iterations. The four baseline intensity
estimates are 0.880, 4.983, 13.182, and 28.059, which are well separated,
representing low, moderate, high, and extremely high components,
respectively. The posterior median of the surface of the
baseline intensity over the study plot, as well as the 2.5\% and
97.5\% percentiles, are shown in Figure~\ref{fig:Realdata_intensity}.
Clearly, after accounting for the available covariates, there are
missing covariates that could have helped to explain the distribution
of the tree species, but they are now captured by the nonparametric
baseline intensity.

Table~\ref{tab:realdata_est} summarizes the posterior mean, standard
deviation, and the 95\% highest posterior density (HPD) credible
intervals of the regression coefficients. Also reported is the posterior
probability of $\gamma_i = 0$ for each covariate $i$, which was used
to decide whether the corresponding covariate is important. The
covariates that are selected are marked with
a star symbol ``*''. The HPD intervals for those covariates that are
not selected either cover 0, or are very close to 0. The tree species
appears to prefer places with higher elevation and steeper slope,
which agrees with the results in~\cite{thurman2014variable}. 
More occurrence of the species is associated with higher
concentrations of Al and Ca, and lower concentrations of Fe, K, P, and
Zn in the soil. These factors' influences on this specific tree species
have not been measured quantitatively before using the BCI dataset.

\begin{table}
  \caption{Posterior means, standard errors (SE), and the 95\% HPD
    credible intervals for the regression coefficients in the analysis
    of the spatial point pattern of \emph{Beilschmiedia pendula} in
    the BCI data. The symbol ``*'' is used to label those significant
    covariates.}
  \label{tab:realdata_est}
  \centering
  \begin{tabular}{ccccc}
    \toprule
Covariate & Posterior $\Pr(\gamma = 1)$ & Estimate & SE & 95\% HPD Credible Interval\\
    \midrule
    elevation & 1.000* & $\phantom{-}$0.694 & 0.065 & ($\phantom{-}$0.575, $\phantom{-}$0.822)\\
    slope & 1.000* & $\phantom{-}$0.651 & 0.039 & ($\phantom{-}$0.570,
                                                  $\phantom{-}$0.727)\\
    pH & 0.015& $\phantom{-}$0.063 & 0.052 &($-0.044$, $\phantom{-}$0.156)\\
    Al & 0.959*   & $\phantom{-}$0.503 & 0.089 & ($\phantom{-}$0.323,
                                        $\phantom{-}$0.671)\\
    B & 0.014 & $-0.045$ & 0.067 & ($-0.171$, $\phantom{-}$0.085)\\ 
    Ca & 1.000* & $\phantom{-}$0.991 & 0.110 & ($\phantom{-}$0.765,
                                        $\phantom{-}$1.203)\\
    Cu & 0.020 & $\phantom{-}$0.073 & 0.068 & ($-0.061$, $\phantom{-}$0.198) \\
    Fe & 0.543* & $-0.306$           & 0.110 & ($-0.501$, $-0.099$)\\
    K & 0.817*  & $-0.431$           & 0.121 & ($-0.624$, $-0.180$)\\
    Mg & 0.032 & $\phantom{-}$0.078 & 0.075 & ($-0.067$, $\phantom{-}$0.222) \\
    Mn & 0.117 & $\phantom{-}$0.187 & 0.064 & ($\phantom{-}$0.064, $\phantom{-}$0.309)\\
    P & 0.915* & $-0.422$           & 0.070 & ($-0.559$, $-0.282$)\\
    Zn & 0.678* & $-0.364$           & 0.105 & ($-0.542$, $-0.155$)\\
    N & 0.019 & $-0.097$ & 0.052 & ($-0.197$, $\phantom{-}$0.005) \\
    N.min & 0.021 & $\phantom{-}$0.063 & 0.061 & ($-0.067$, $\phantom{-}$0.173)\\
    \bottomrule
  \end{tabular}
\end{table}

We also fitted other three competing models in the simulation study and
compared the MSE of the models calculated on the $50\times 25$ grid.
In the spline-NHPP model, cubic B-spline basis were used with three
knots for $x$ and one knot for $y$, since the range of $x$ is twice as
long as that of $y$, resulting 7 and 5 degrees for $x$ and $y$,
respectively, in the tensor product spline basis construction.
The proposed method gives an MSE of only 6.00. The MSE from the 
const-NHPP, spline-NHPP, and LGCP models are 26.69, 19.98 and 26.70,
respectively. The improvement made by the proposed method is obvious.

\section{Discussion}
\label{sec:discussion}

Explaining the spatial heterogeneity of point patterns is challenging
when important covariates are not observed. The proposed
semiparametric NHPP model captures the heterogeneity unexplained by
observed covariates with a spatially varying baseline intensity. The
baseline intensity surface is of a flexible form of piecewise constant
on a grid partition of the study region, with a PCRP prior that
prevents overly small components often seen when a CRP prior is
imposed instead. The methodology is particularly useful when the
baseline intensity surface lacks smoothness as in the case of missing
covariates. The fitted number of components of the piecewise constant
baseline depends on the power $r$ of the PCRP. The selection of $r$
thorough the BITC specifically designed for this setting seems to be
more effective in the simulation study than the LPML and DIC.

A few topics beyond the scope of this paper merits further
investigation. When the baseline intensity is deemed to be spatially
contiguous, imposing spatial contiguity on the piecewise constant
intensity surface \citep{li2019spatial} may lead to more
efficient estimator of the surface. Some applications may have large
areas containing no events, in which case, including zero-inflated
structures \citep{lambert1992zero} in spatial point process models may
improve the fitting and avoid unidentifiably close-to-zero
intensities. Finally, selection of $r$ for the PCRP prior by the BITC
needs to be evaluated in more general settings. A tuning-free
strategy, for example, through a hyper prior on $r$, would be
desirable for practice.

\section*{Acknowledgements}
The authors thank Drs. Dipak Dey and Yishu Xue for their comments and
suggestions. GH's research was supported by Dean’s office
of the College of Liberal Arts and Sciences at the University of Connecticut.
The BCI forest dynamics research project was founded by S.P. Hubbell
and R.B. Foster and is now managed by R. Condit, S. Lao, and R. Perez
under the Center for Tropical Forest Science and the Smithsonian
Tropical Research in Panama. Numerous organizations have provided
funding, principally the U.S. National Science Foundation, and
hundreds of field workers have contributed.

\section*{Supporting Information}
\subsection*{Appendix A}
This section shows the derivation of full conditionals needed for the
MCMC. For the full conditional distribution of
$\lambda_{0, i}$, we only need to focus on those data points
that are in the $i$th component since the likelihood for the data
points in other components does not involve $\lambda_{0, i}$.
That is, we only need to focus on
those grid boxes $\{A_j\}$'s such that $z_j = i$.
The full conditional density of $\lambda_{0, i}$,
$i = 1, \ldots, n$, is
\begin{align}\label{eq:post_lambda}
  \begin{split}
    q(\lambda_{0, i} \mid \mathbf{S}, \bm{\beta},
    \bm{\gamma}, \bm{z}, \bm{\lambda}_{0, -i})
    &\propto
    \frac{\prod_{\ell:\bm{s}_\ell \in A_j,  z_j = i}\lambda(\bm{s}_\ell)}
    {\exp(\int_{\bigcup_{j:z_j = i}A_j}\lambda(\bm{s}) \dd \bm{s})}
    \lambda_{0,i}^{a-1}\exp\left(-b\lambda_{0, i}\right)\\
    &=
    \frac{\prod_{\ell:\bm{s}_\ell \in A_j, z_j = i} \lambda_{0, i}
      \exp(\mathbf{X}^\top(\bm{s}_\ell)\bm{\beta})}
    {\exp\left(\lambda_{0, i} \int_{\bigcup_{j:z_j =  i}A_j}
      \exp\left(\mathbf{X}^\top(\bm{s})\bm{\beta}\right)\dd \bm{s}\right)}
    \lambda_{0,i}^{a-1}\exp\left(-b\lambda_{0, i}\right)\\
    &\propto
    \lambda_{0, i}^{N_i+a-1}
    \exp\left(-\left(b + \sum_{j:z_j = i}
        \Lambda_j(\bm{\beta})\right)\lambda_{0, i}\right),
  \end{split}
\end{align}
which is the density of
$\mbox{Gamma}\left(N_i+a,
  b+\sum_{j:z_j = i}\Lambda_j(\bm{\beta})\right)$.

The full conditional mass function of $\gamma_i$,
$i = 1, \ldots, p$, is
\begin{equation}\label{eq:post_gamma}
  \begin{split}
    q(\gamma_i \mid
    \mathbf{S}, \bm{\beta}, \bm{\gamma}_{-i}, \bm{z}, \bm{\lambda}_0)
    &\propto 0.5^{\gamma_i} 0.5 ^{1-\gamma_i}
    \phi^{\gamma_i} (\beta_i | 100)
    \phi^{1 - \gamma_i} (\beta_i | 0.01)\\
    &= \big(0.5 \phi(\beta_i | 100) \big)^{\gamma_i}
    \big(0.5 \phi(\beta_i | 0.01) \big)^{1 - \gamma_i},
  \end{split}
\end{equation}
which is Bernoulli with rate parameter
\[
  \frac{0.5 \phi(\beta_i | 100)}
  {0.5 \phi(\beta_i | 100) + 0.5\phi(\beta_i | 0.01)}.
\]

The full conditional density of $\beta_i$, $i = 1, \ldots, p$, is
\begin{equation}\label{eq:post_beta}
  \begin{split}
    q(\beta_i \mid
    \mathbf{S}, \bm{\beta}_{-i}, \bm{\gamma}, \bm{z}, \bm{\lambda}_0)
    &\propto 
    \phi^{1-\gamma_i}\left(\beta_i  | 0.01\right)
    \phi^{\gamma_i}\left(\beta_i | 100\right) \\
    & \phantom{=} \times \prod_{i=1}^n \lambda_{0, z_i}^{m_i}
    \exp\left(\sum_{j:\bm{s}_j \in A_i}
      \mathbf{X}^\top(\bm{s}_j)\bm{\beta} - \lambda_{0,
        z_i}\Lambda_i(\bm{\beta})\right).
  \end{split}
\end{equation}
This is not a standard distribution. Sampling from it can be done
with a Metropolis--Hasting algorithm with a normal proposal
distribution centered at the value of the current iteration with a
variance parameter tuned to achieve desired acceptance rate.

The full conditional distribution of $z_i$, $i = 1, \ldots, n$,
is contingent on whether the $i$th grid box goes to an existing
component or a new one \citep{neal2000markov}.
The full conditional probability that grid box $A_i$
belongs to an existing component $c$, i.e.,
$\exists j \ne i,\, z_j = c$, is
\begin{align}\label{eq:post_zexist}
  \begin{split}
    \Pr(z_i = c \mid
    \mathbf{S}, \bm{z}_{-i}, \bm{\lambda}_0, \bm{\beta})
    &\propto
    \frac{n_{-i,  c}^r}{\sum_{j = 1}^k n_j^r-1+\alpha}
    \frac{\lambda_{0, c}^{m_i}
      \prod_{j:\bm{s}_j \in A_i}
      \exp(\mathbf{X}^\top(\bm{s}_j)  \bm{\beta})}
    {\exp(\lambda_{0, c}\Lambda_i(\bm{\beta}))}\\
    &=
    \frac{n_{-i, c}^r}{\sum_{j = 1}^k n_j^r-1+\alpha}
    \lambda_{0, c}^{m_i}
    \exp\left(
      \sum_{j:\bm{s}_j \in A_i} \mathbf{X}^\top(\bm{s}_j)\bm{\beta}
      - \lambda_{0, c}\Lambda_i(\bm{\beta})\right).
  \end{split}
\end{align}
The full conditional probability that $A_i$ belongs to a new
component, i.e., $\forall j\ne i,\, z_j \ne c$, is
\begin{align}\label{eq:post_znew}
  \begin{split}
    &\phantom{ = } \Pr(z_i = c\mid \mathbf{S}, \bm{z}_{-i},
    \bm{\lambda}_0, \bm{\beta}) \\
    & \propto
    \frac{\alpha}{\sum_{j = 1}^k n_j^r-1+\alpha}
    \int
    \frac{\lambda_{0,  c}^{m_i}
      \prod_{j:\bm{s}_j \in A_i}\exp(\mathbf{X}^\top(\bm{s}_j)
      \bm{\beta})}
    {\exp\left(\lambda_{0, c}\Lambda_i(\bm{\beta})\right)}
    \frac{b^a}{\Gamma(a)}\lambda_{0,c}^{a-1} e^{-b\lambda_{0,c}}
    \dd \lambda_{0, c}\\
    &=
    \frac{\alpha}{\sum_{j = 1}^k n_j^r-1+\alpha}
    \left(\prod_{j:\bm{s}_j \in A_i}
      \exp(\mathbf{X}^\top(\bm{s}_j) \bm{\beta})\right)
    \frac{b^a}{\Gamma(a)}
    \int
    \lambda_{0, c}^{m_i+a-1}e^{-(b+\Lambda_i(\bm{\beta}))\lambda_{0, c}} \dd
    \lambda_{0, c}\\
    &= \frac{\alpha b^a \Gamma(m_i+a)}
    {(\sum_{j = 1}^k n_j^r-1+\alpha)
      (b+\Lambda_i(\bm{\beta}))^{m_i+a} \Gamma(a)}
    \exp\left(\sum_{j:\bm{s}_j \in  A_i}
      \mathbf{X}^\top(\bm{s}_j)\bm{\beta}\right).
  \end{split}
\end{align}
Combining~\eqref{eq:post_zexist} and~\eqref{eq:post_znew} gives the
full conditional distribution of $z_i$ in the main text.

\subsection*{Appendix B}
\begin{figure}[tbp]
  \centering
  \subcaptionbox{Setting 1}{
    \includegraphics[width=\textwidth]{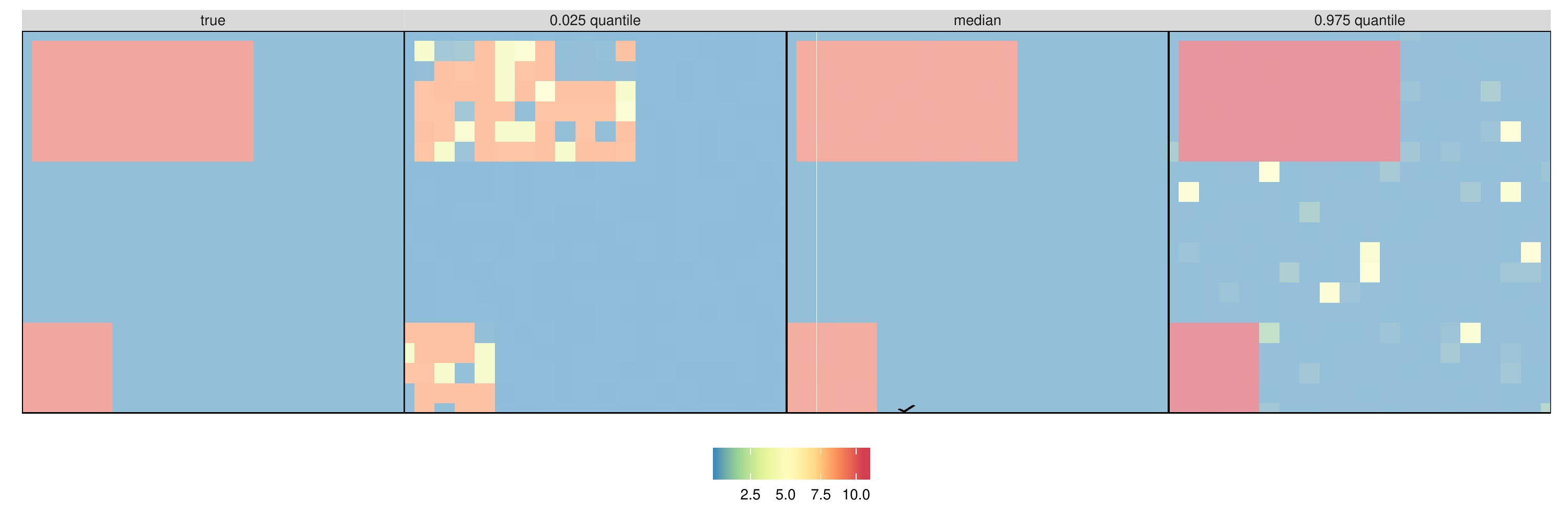}}
  \subcaptionbox{Setting 2}{
    \includegraphics[width=\textwidth]{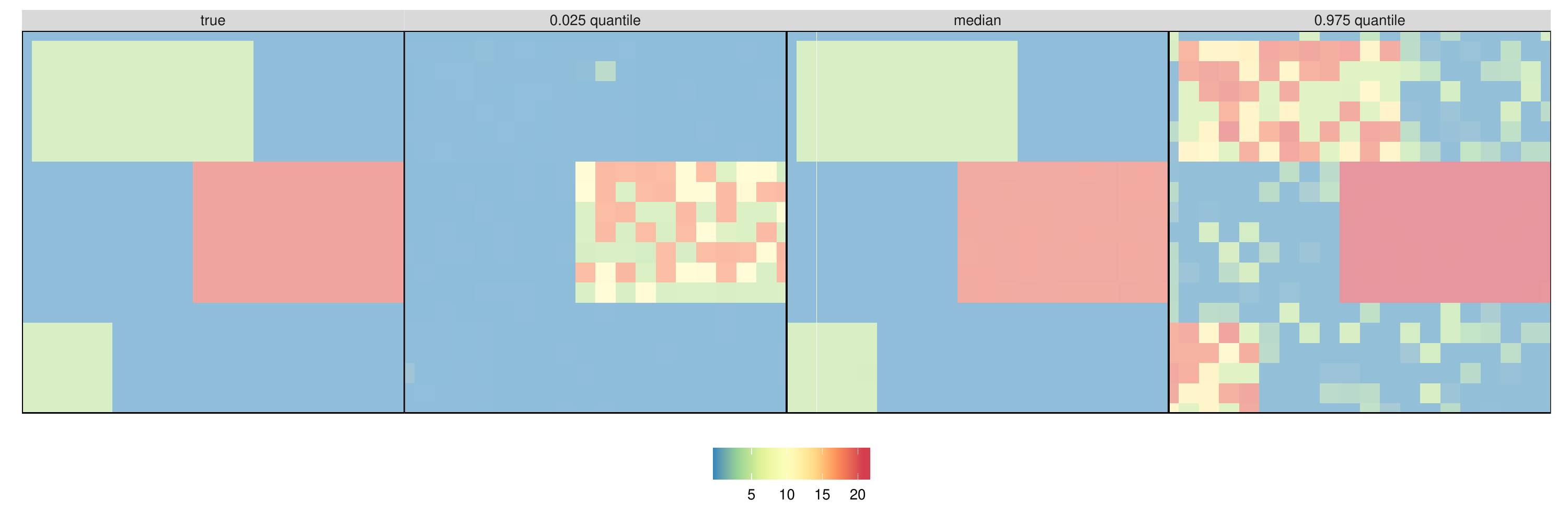}}
  \caption{Simulation configurations for baseline intensity, with
    fitted baseline intensity surfaces. Median and quantiles are
    calculated out of 100 replicates.}
  \label{wfig:baseline}
\end{figure}

This section includes supporting materials for simulation study. There
are two different settings for baseline intensity in simulation
study. The heatmaps showing the configuration of baseline intensity
surfaces, and the fitted surfaces corresponding to the $2.5\%$ quantile,
median and $97.5\%$ quantile of 100 replicates are displayed in
Figure~\ref{wfig:baseline}.

\begin{figure}[tbp]
  \centering
  \includegraphics[width = \textwidth]{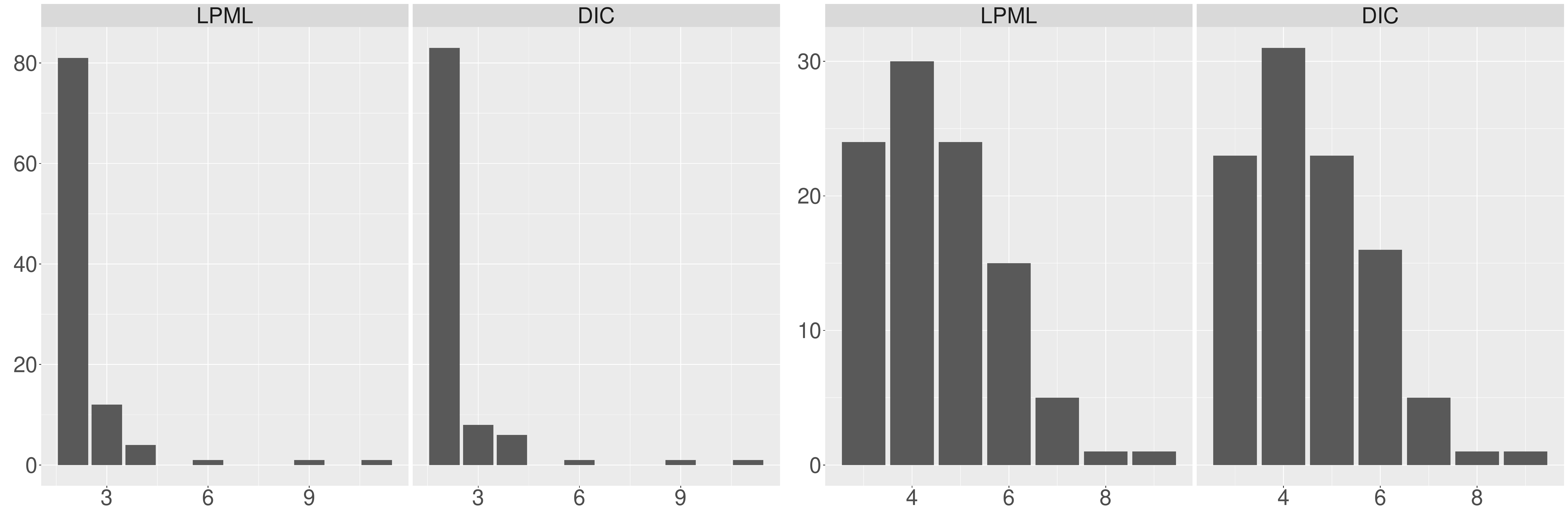}
  \caption{Histogram of $\hat{K}$ chosen by LPML and DIC over 100
    replicates. Results of setting 1 and 2 are on left and right
    panel, respectively.}
  \label{wfig:K_hist2}
\end{figure}

The performances of LPML and DIC to select power $r$ are not
good. The have satisfying results under setting 1, where most of the
time they give correct estimation for $K = 2$, but under setting 2,
when the problem is more difficult, both of them fail to give accurate
estimate for $K$. The histgrams of $\hat{K}$ chosen by LPML and DIC
over 100 replicates are shown in Figure~\ref{wfig:K_hist2}.

\subsection*{Appendix C}
\begin{figure}[tbp]
  \centering
  \includegraphics[width=\textwidth]{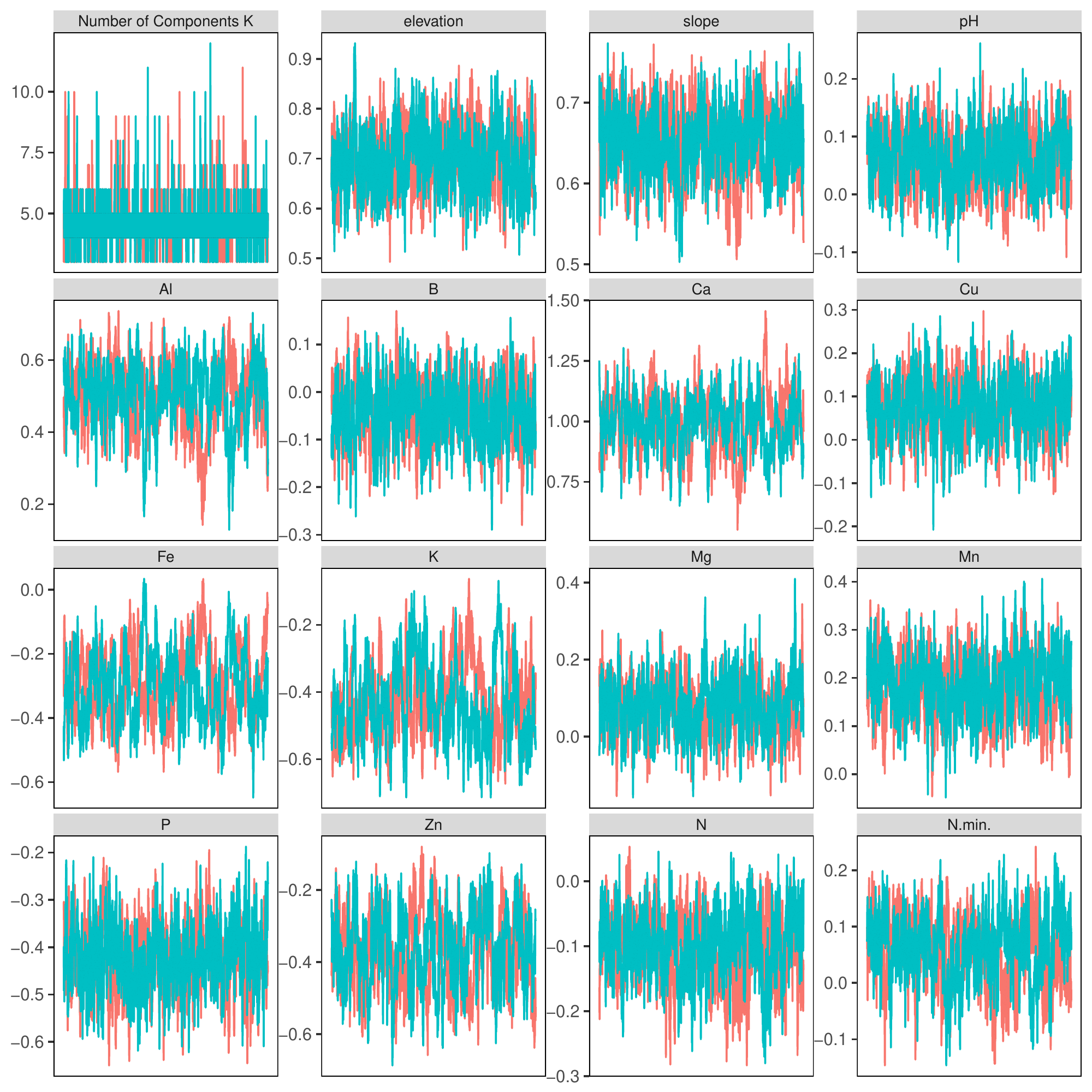}
  \caption{BCI data analysis: trace plots of $\bm{\beta}$ and $K$
    after burnin and thinning.}
  \label{wfig:trace}
\end{figure}

The trace plots of $\bm{\beta}$ and $K$ for BCI data fitting results
after burnin and thinning are shown in Figure~\ref{wfig:trace}. The
convergence of each element in $\bm{\beta}$ is satisfying. The trace
plot of $K$ shows good convergence of the grouping
process~\citep{wang2013bayesian}. Since Dahl's method requires that
the chain has converged, we first checked the convergence using trace
plot of $K$. The estimation for $\hat{\bm{z}}$ was then calculated
using Dahl's method. It was then used in place of the truth to
construct the trace plot of RI, which was used to further check the
convergence of $\bm{z}$.

\label{lastpage}

\bibliographystyle{chicago}  
\bibliography{ms}

\end{document}